\documentclass[prb,superscriptaddress,showpacs,papersize=a4paper,twocolumn]{revtex4}

\usepackage{caption}
\usepackage{subcaption}
\usepackage{amsmath}
\usepackage{amsfonts}
\usepackage{amssymb}
\usepackage{graphicx}
\usepackage[usenames]{color}
\usepackage{hyperref}

\begin{document}

\title{Admittance of a long diffusive SNS junction}
\author{K. S. Tikhonov}
\email{tikhonov@itp.ac.ru}
\affiliation{L. D. Landau Institute for Theoretical Physics, 117940 Moscow, Russia}
\affiliation{Moscow Institute of Physics and Technology, 141700 Moscow, Russia}
\affiliation{Department of Physics \& Astronomy, Texas A\&M University, College Station,
TX 77843-4242, USA}
\author{M. V. Feigel'man}
\affiliation{L. D. Landau Institute for Theoretical Physics, 117940 Moscow, Russia}
\affiliation{Moscow Institute of Physics and Technology, 141700 Moscow, Russia}

\pacs{74.45.+c, 74.25.N-, 74.40.Gh}

\begin{abstract}
The dynamical properties of hybrid normal metal/superconductor structures
have recently come into research focus both experimentally and
theoretically. Recent experimental studies of the coherent admittance $%
Y(\omega )$ of SNS rings as function of the phase difference $\phi _{0}$ are
still not fully understood. Here we concentrate on the linear response
regime, calculating $Y(\omega )$ by solving Usadel equations, linearised in
electric field. Although partially reproducing previously known results, we
find qualitatively different behaviour in the collisionless regime of $\tau
_{in}^{-1}\ll \omega \lesssim E_{Th}$ and high temperature $T\gg E_{Th}$ and
low temperature $T\lesssim E_{Th}$ near the minigap closing $\phi _{0}\sim
\pi $. We find that the dissipative part $\mbox{Re}Y(\omega )$ peaks when
the minigap closes (at a phase difference of $\pi $) even at high
temperatures, when the equilibrium supercurrent is fully suppressed.
\end{abstract}

\maketitle

\section{Introduction}

\begin{figure}[ptb]
\centering
\begin{subfigure}{.25\textwidth}
\centering
\includegraphics[width=.9\linewidth]{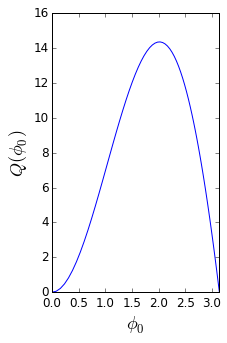}
\label{fig:sub1}
\end{subfigure}%
\begin{subfigure}{.25\textwidth}
\centering
\includegraphics[width=.9\linewidth]{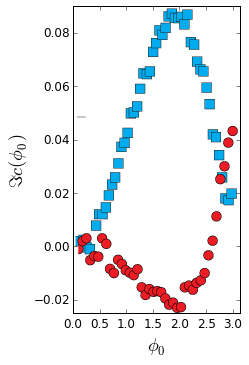}
\label{fig:sub2}
\end{subfigure}
\caption{To the left: function $Q(\protect\phi_{0})$, entering Lempitskii's
prediction for coherent part of the dissipation. To the right: Dissipative
response ($c=\protect\chi/\protect\chi_{0}$) at $T\approx17E_{Th}$, data
from Ref. \onlinecite{dassonneville14}. Blue curve: hydrodynamic regime, $%
\hbar \protect\omega=0.4E_{Th}$, red curve: collision-less regime, $\hbar%
\protect\omega=2E_{Th}$ (arbitrary shifted in $c$ axis.)}
\label{fig:test}
\end{figure}

The superconducting proximity effect on the transport properties of normal
metal/superconductor structures has been thoroughly studied both
theoretically and experimentally. Most of the studies were concentrated on
properties of these systems in equilibrium\cite{pannetier00}. Recently, one
of the most basic quantities, characterizing dynamical properties of such
structures, - the admittance $Y\left( \omega\right) =I\left( \omega\right)
/V\left( \omega\right) $, acquired more attention. It characterizes the
current response $I\left( t\right) =\int\left( d\omega\right) I\left(
\omega\right) e^{-i\omega t}$ to an \emph{ac} voltage $V\left( t\right)
=\int\left( d\omega\right) V\left( \omega\right) e^{-i\omega t}$ in the
linear response regime.

The problem of calculation the current in the tunnelling (SIS) junction has
been solved long ago for arbitrary time-dependent voltage $V(t)$\cite%
{larkin67}. The phase dynamics of such a junction, coupled to the
electromagnetic environment can usually be described by RSJC model\cite%
{likharev85}. The same problem for a superconductor - normal metal -
superconductor (SNS) junction is much more complicated, since the \emph{ac}
dynamics of the phase interferes here with the dynamics of the electrons in
the normal metal. Additionally, multiple Andreev reflections are very
important in such junctions, producing highly non-trivial energy
distribution of the electrons in the wire\cite{zaitsev98, bezuglyi00,
cuevas06}, but they are not essential in the regime of small voltage which
we concentrate on.

As follows, superconducting proximity effect causes $Y$ to be different from 
$Y_{N}=1/R_{N}$, the admittance of the wire in the normal state. Due to the
Josephson relation $\dot{\phi}=2eV/\hbar$, the admittance can be related to
the linear susceptibility of the junction with respect to the oscillating
superconducting phase difference. In the geometry of an SNS ring, the phase
difference $\phi$ $=-2\pi\frac{\Phi}{\Phi_{0}}$ (where $\Phi_{0}=h/2e$ is
the flux quantum) is controlled by magnetic flux $\Phi$ penetrating this
ring. The corresponding response function $\chi\left( \omega\right) =\frac{%
\delta I}{\delta\Phi}$ can be directly measured, and is related to $Y\left(
\omega\right) $ as follows: 
\begin{equation}
\chi\left( \omega\right) =i\omega Y\left( \omega\right) .   \label{chi}
\end{equation}
In practice, the measurement of $\chi$ can be conducted by exposing the SNS\
ring to a weak magnetic field $B\left( t\right) =B_{0}+B_{osc}e^{-i\omega t}$%
. While $B_{0}$ fixes the stationary part of the superconducting phase
difference along the normal wire $\phi_{0}=-2\pi \frac{\Phi}{\Phi_{0}}$, $%
B_{osc}$ generates an e.m.f. $\mathcal{E}\left( t\right) =-\frac{1}{c}\frac{%
d\Phi_{osc}}{dt}$, generating an $ac$ electric current.

In the static limit ($\omega\rightarrow0$), $\mathcal{E}\left( t\right) $ is
absent and the equilibrium response function $\chi\left( 0\right) $ is
recovered: 
\begin{equation}
\chi\left( \omega=0\right) =-\frac{2\pi}{\Phi_{0}}\partial_{\phi}I_{S}\left(
\phi_{0}\right) ,   \label{adia}
\end{equation}
where $I_{S}\left( \phi_{0}\right) $ stands for the current-phase relation
of the junction in equilibrium. At finite frequency, the effect of $\mathcal{%
E}\left( t\right) $ is to both modify the non-dissipative response and to
generate dissipation in the normal wire.

It can be expected that there should exist a limit, in which the admittance $%
Y$ of the junction equals the admittance of two $SN$ junctions connected in
series. In this \emph{incoherent} limit, \thinspace$Y$ is $\phi_{0}$%
-independent. As we will demonstrate shortly, it is achieved only when
frequency is large $\omega\gg E_{Th}$, that is, the admittance has a
significant \emph{coherent} contribution even at very large temperature $%
T\gg E_{Th}$, but moderate frequency $\omega\lesssim E_{Th}$. This
contribution will be the focus of our discussion. Another interesting point
is that by measuring the response of the wire to $\mathcal{E}$ as a function
of frequency and $dc$ phase difference $\phi_{0}$\cite{chiodi11,
dassonneville13}, one infers the dynamical properties of the Andreev levels
in the junction through their effect on the conductive properties of the
normal wire. This effect is thus very sensitive to inelastic processes in
the wire and can be used as a specific probe.

From the theoretical side, the first study of the coherent contribution to
the impedance of an SNS bridge was performed by Lempitskii\cite{lempitskii83}%
. He considered the long junction limit ($E_{Th}\ll\Delta$) at high
temperature ($T\gg E_{Th}$), biased in the \emph{adiabatic} regime ($%
\hbar\omega\ll E_{Th}=\hbar D/L^{2}$), and obtained the following result:%
\begin{equation}
Y\left( \omega\right) =Y_{N}\frac{E_{Th}}{T}\frac{E_{Th}/\hbar}{\tau
_{in}^{-1}-i\omega}Q\left( \phi_{0}\right) ,   \label{lemp}
\end{equation}
with $\tau_{in}$ staying for the inelastic relaxation time and universal
function $Q\left( \phi_{0}\right) $ evaluated numerically. This function has
recently been recalculated \cite{virtanen11} with better precision, see Fig. %
\ref{fig:test} for the result.

Lempitskii's effect results from supercurrent-enhanced non-equilibrium
population of Andreev levels in the wire. The most fascinating result is
that this non-equilibrium population causes the coherent part of $Y$ to
decay (at given $\tau_{in}$) slowly, as $E_{Th}/T,$ at $T\gg E_{Th}$,
whereas the equilibrium supercurrent decays exponentially, as $%
\propto\exp\left( -L/L_{T}\right) $. This non-equilibrium enhancement of the
superconducting correlations recalls the well known effect of the microwaves
enhancement of superconductivity, the phenomena, known as the Dayem-Wyatt
effect\cite{dayem67, wyatt66,eliashberg70}, which is observed in
microbridges, thin films and stripes\cite{kommers77, dahlberg79, hall80,
klapwijk77, hamer84}. Similar effect exist in the hybrid structures\cite%
{notarys73, warlaumont79}, but their physics is enriched by existence of two
different time scales: time of diffusion along the normal part $\tau_{D}$
and inelastic scattering rate $\tau_{in}$, as was clearly demonstrated
recently \cite{chiodi09}. In our work, we concentrate on how this rich
physics shows itself in the linear response function $Y$.

Since Lempitskii's work, there was not much theoretical activity on the
coherent contribution to $Y$ with notable exceptions provided by\cite%
{zhou97,argaman99}. However, the recent experiments motivated a series of
theoretical studies \cite{virtanen10, virtanen11, ferrier13}. In particular,
extensive numerical work\cite{virtanen10, virtanen11}, supported by
qualitative analytical treatment, was devoted to study $Y$ in a wide range
of temperatures and frequencies.

Detailed comparison of the existent theoretical predictions to the
experimental results was performed in Ref. \onlinecite{dassonneville14}. It
was found that the non-dissipative response $\mbox{Im} Y$ of the junction
can be well understood on the basis of Lempitskii's theory for all moderate
frequencies: $\hbar\omega\lesssim E_{Th}$ (In that experiment,$~E_{Th}=71mK$%
, corresponding to the frequency $f_{Th}=1.5GHz$). Interestingly,
experimental results demonstrate that it is possible to follow the response
function while it crosses over from hydrodynamic ($\omega\tau_{in}\ll1$) to
collision-less ($\omega\tau_{in}\gg1$) regime, and extract inelastic
scattering rate $\tau_{in}^{-1}$ as a function of temperature. At the
highest temperature studied, $T\approx1.2K$ it was found that $%
\tau_{in}\approx2.5\tau_{D}$. Interestingly, the scattering rate, found in
this experiment, demonstrates unusual temperature dependence, $%
\tau_{in}^{-1}\propto T^{2}$. We are not aware of any physical mechanism
which can lead to such a dependence in a normal gold wire and believe that
this power law is specific to the wire in the conditions of the proximity
effect. We expect that it is related to the strong modification of the
electronic spectrum in the wire by the superconducting contacts, which
should influence electron-electron scattering processes - the effect which
certainly deserves future studies.

Experimental results for hydrodynamic and collision-less regimes are
presented by Fig. \ref{fig:test}. In these figures, phase susceptibility is
measured in natural units: 
\begin{equation}
c=\chi/\chi_{0},   \label{cdim}
\end{equation}
with $\chi_{0}=Y_{N}E_{Th}/\hbar$. Theoretical prediction for this quantity,
obtained from Eqs (\ref{chi}) and (\ref{lemp}), gives:%
\begin{equation}
c^{th}=\frac{E_{Th}}{T}\frac{i\omega}{\tau_{in}^{-1}-i\omega}Q(\phi_{0}). 
\label{cY}
\end{equation}
Fig. \ref{fig:test} illustrates one of the most important experimental
observations: while at low frequency $\mbox{Im} c^{exp}$ fits well with
Lempitskii's result (see Fig. \ref{fig:test} for $Q\left( \phi_{0}\right) $%
), at higher frequency ($\omega\tau_{in}>1$) the dissipative response has a
very different shape as a function of \emph{dc} phase bias $\phi_{0}$.
Recall that Eq. (\ref{cY}) results from the adiabatic calculation, which
assumes $\omega\ll E_{Th}$. One may hope that the full numerical calculation
(the one not relying on the expansion in $\omega/E_{Th}$) in this regime can
describe the experimental results. Such a calculation, performed in the Ref. %
\onlinecite{virtanen11} implies that the peak in $\mbox{Im} c^{exp}$ at
minigap closing $\phi_{0}=\pi$ should be absent at $T\gg E_{Th}$, in clear
contradiction with experiment.

This contradiction motivated a linear response analysis on the basis of BdG
equations \cite{ferrier13}. The results of the latter study seem to indicate
qualitatively the presence of a maximum at $\phi_{0}=\pi$. From the
theoretical side, it is clear that Lempitskii's prediction concerns
deviation from equilibrium of the Andreev pairs only, but quasi-particle
excitations in the normal region can also be relevant, especially at not too
low frequency. While it is clear that dissipation due to quasi-particles,
excited by electric field, should be sharply peaked at $\phi_{0}=\pi$ at low
temperatures $T\ll E_{Th}$, the fate of this peak at high temperature is not
obvious \emph{a priori}. As we mentioned above, it was predicted in Ref. %
\onlinecite{virtanen11}, that at high temperature this peak should
disappear. Our goal is to reconsider this problem and to resolve the
apparent contradiction between numerical results and experimental data in
this regime.

We start with a simple observation: for the dissipative response the
condition for validity of the adiabatic approximation is more stringent than
for the non-dissipative one. Since the adiabatic contribution to $\mbox{Im} c
$ decreases for $\omega\gtrsim\tau_{in}^{-1}$, the non-adiabatic
(proportional to $\frac{\hbar\omega}{E_{Th}}$) correction to Eq (\ref{lemp})
becomes essential already at $\hbar\omega\sim\sqrt{\hbar E_{Th}/\tau_{in}}%
\ll E_{Th}$. As we will show, the terms of the order of $\hbar\omega/E_{Th}$
result from the charge imbalance (induced by the \emph{ac} electric field)
and lead to the enhancement of dissipation at $\phi_{0}\approx\pm\pi$.

Our approach is based on the Usadel equation, expanded to the first order in
the electric field, without assuming smallness of the proximity effect (in
particular, we take into account all non-perturbative effects, such as the
minigap). Although it is impossible to get a response function $Y\left(
\omega,\phi_{0}\right) $ in closed form even in the simplest limiting cases,
we go as far as possible analytically, resorting to numerical calculation
only at the latest stage, which makes our calculation more controllable than
fully numerical solution of the time-dependent Usadel equation.

\section{Usadel equation and linear response}

\subsection{General equations.}

In what follows we make several additional simplifying assumptions: i) we
treat the system as quasi one-dimensional, ii) we treat electron-electron
interaction in the wire in the relaxation time approximation, neglecting
possible energy and position dependence of the relaxation time as well as
its modification by the proximity effect, and iii) we assume that $\Delta
/E_{Th}\gg1$. We measure the energy in units of $E_{Th}$ and length in units
of $L$. Our starting point is Usadel equation ($e=-\left\vert e\right\vert $%
) in the presence of electric field. Due to gauge invariance, we are allowed
to use scalar potential $\varphi$ instead of vector potential $\mathbf{A}$
to define an electric field in our quasi-one-dimensional normal wire: $%
E=-\nabla\varphi$. Then Usadel equation acquires the form:

\begin{equation}
\partial_{x}\left( \check{g}\cdot\partial_{x}\check{g}\right) +i\left[
\epsilon\hat{\tau}^{3},\check{g}\right] +ie\left[ \varphi,\hat{g}\right]
=I_{St}\left[ \check{g}\right] .   \label{Usadel}
\end{equation}
In this equation $\check{g}\left( x,t,t^{\prime}\right) $ is isotropic part
of quasi-classical Keldysh Green function, which is a matrix in the
Nambu-Gorkov space. In terms of this function, the electric current can be
expressed as follows ($S$ stands for the area's wire):$\ $%
\begin{equation}
I=\frac{\pi\sigma_{N}S}{4e}\mbox{tr}\left( \hat{\tau}^{3}\hat{\jmath}%
^{K}(t,t)\right) ,   \label{curr}
\end{equation}
where $\check{j}=\check{g}\cdot\nabla\check{g}.$

Neglecting spatial gradients in the superconducting reservoirs, we write for
the Green function there: 
\begin{equation}
\check{g}_{S}=\check{S}_{\phi}\cdot\check{g}_{eq}\cdot\check{S}_{\phi}^{+}, 
\label{gs}
\end{equation}
with 
\begin{equation}
\check{S}\left( t,t^{\prime}\right) =\delta\left( t-t^{\prime}\right) e^{i%
\hat{\tau}_{3}e\int^{t}\varphi\left( \tau\right) d\tau}.   \label{gauge}
\end{equation}
Here $\check{g}_{eq}$ is the equilibrium BCS Green function.

The Usadel equation (\ref{Usadel}) includes spectral and kinetic equations
which may be obtained with the use of conventional parametrization $\hat {g}%
^{K}=\hat{g}^{R}\cdot\hat{H}-\hat{H}\cdot\hat{g}^{A}$, where $\hat{H}$ is a
diagonal matrix of distribution functions in the Nambu-Gorkov space. In
equilibrium, distribution function equals $\hat{H}=h\left( \epsilon\right) 
\hat{\tau}_{0},$ with $h\left( \epsilon\right) =\tanh\frac{\epsilon}{2T}.$
For retarded Green function the following parametrization is appropriate:

\begin{equation}
\hat{g}_{eq}^{R}\left( \epsilon,x\right) =\left( 
\begin{array}{cc}
G & F \\ 
\bar{F} & -G%
\end{array}
\right)   \label{g0}
\end{equation}
where $G=\cosh\theta,$ $F=\sinh\theta e^{i\phi},~\bar{F}=-\sinh\theta
e^{-i\phi}$. In this parametrization, the spectral angle $\theta$ satisfies:

\begin{equation}
\partial_{x}^{2}\theta+\left( 2i\epsilon-\tau_{in}^{-1}\right) \sinh
\theta+J^{2}\frac{\cosh\theta}{\sinh^{3}\theta}=0,   \label{UsadelTheta}
\end{equation}
where $J\equiv J(\epsilon)=i\sinh^{2}\theta\partial_{x}\phi$ is the spectral
supercurrent, which is an integral of motion: $\partial_{x}J=0$ and we
employed the relaxation time approximation. The boundary conditions for $\phi
$ and $\theta$ are fixed by the BCS functions.

It is not feasible to write down the solutions of Eq (\ref{UsadelTheta}) in
a closed analytical form. However, the properties of the solutions are well
known and numerical approaches to it are well developed. In order to obtain
the solutions, we use publicly available solver, developed by P. Virtanen
and T. Heikkila and described in Ref \onlinecite{usnum}. It provides Green
function in the Ricatti parametrization, which is related to the
trigonometric parametrization by means of the equations presented in the
Appendix \ref{AA}.

Once the unperturbed solution is found, the effects of the weak electric
field can be discussed. In the presence of oscillating electric potential $%
\varphi$, the Green function becomes time dependent: $\check{g}=\check{g}%
_{eq}+\delta\check{g}$. The effect of the electric field is twofold. First,
it imposes time dependence on the phases of the order parameters in the
superconducting contacts, see Eq. (\ref{gs}). This modifies the spectrum of
the energy levels in the junction through corrections to retarded and
advanced Green functions. Second, it induces inter-level transitions with
energy transfer $\omega,$ changing the populations of these levels through
corrections to the distribution function. Contributions of these two types
of corrections to the electric current behave very differently at high
temperatures: the former decay exponentially $\propto\exp\left(
-L/L_{T}\right) ,$ while the latter decreases as a power-law with increasing
the temperature.

\subsection{Kinetic corrections.}

Let us start with a discussion of the correction to the generalized
distribution function $\delta\hat{H}$. It can be chosen diagonal in the
particle-hole space:

\begin{equation*}
\delta\hat{H}\left( \epsilon,\epsilon^{\prime},x\right) =\left[ h_{L}\left(
\epsilon,x\right) \hat{\tau}_{0}+h_{T}\left( \epsilon,x\right) \hat{\tau}_{3}%
\right] \delta\left( \epsilon-\epsilon^{\prime}-\omega\right) \text{.}
\end{equation*}
In the contacts, the transversal distribution function $h_{T}$ is driven out
of equilibrium by the time-dependent voltage:

\begin{equation}
h_{T}\left( \epsilon,x=0,1\right) =h_{T,0}\left( \epsilon,x=0,1\right) ,
\end{equation}
The function $h_{T}(\epsilon,x)$ describes charge imbalance that is induced
in the N region due to oscillating electric field.

The longitudinal distribution function $h_{L}(\epsilon,x)$ describes all
deviations from the equilibirum Fermi distribution function $h(\epsilon)$,
which are related with non-equilibrium in energy distribution, but without
any charge imbalance. $h_{L}(\epsilon,x)$ remains unperturbed within the
linear response regime strictly at the boundaries with both superconductors: 
\begin{equation}
h_{L}\left( \epsilon,x=0,1\right) =0,   \label{bcl}
\end{equation}
however it varies sharply within a short distance near these boundaries, as
will be discussed below.

In the wire, $h_{L,T}$ are governed by conservation laws of energy and
charge currents:

\begin{gather}
\partial_{x}j_{L}+N\left( i\omega-\tau_{in}^{-1}\right) h_{L}=0,  \label{kl}
\\
\partial_{x}j_{T}+N\left( i\omega-\tau_{in}^{-1}\right) \left[ h_{T}-h_{T0}%
\right] =0,   \label{kt}
\end{gather}
where relaxation time approximation is employed. Here%
\begin{equation*}
h_{T,0}\left( \epsilon,x\right) =e\varphi\left( x\right) \frac{h\left(
\epsilon-\omega\right) -h\left( \epsilon\right) }{\omega}, 
\end{equation*}
and $h(\epsilon)=\tanh(\epsilon/2T)$ is equilibrium Fermi distribution
function. Spatial distribution of the electric potential, $\varphi\left(
x\right) $ has to be found from the Poisson equation:%
\begin{equation}
\Delta\varphi=-\rho,   \label{pois}
\end{equation}
taking into account the fact that the oscillating voltage drop along the
wire $Ve^{-i\omega t}=\varphi\left( 0\right) -\varphi\left( 1\right) $ is
fixed by the applied \textit{ac} phase modulation. In general, this gives a
complicated coupled system of equations (\ref{Usadel}) and (\ref{pois})
which can be solved iteratively. In general, we find that for all
frequencies of interest, the corrections to%
\begin{equation}
\varphi\left( x\right) =\varphi\left( 0\right) -Vxe^{-i\omega t}, 
\label{simpl}
\end{equation}
which result from the charge redistribution in the wire do not lead to
noticable modification of the coherent part of the admittance and we neglect
them in what follows.

The energy current in Eq (\ref{kl}) reads: 
\begin{equation}
j_{L}=D_{L}\partial_{x}h_{L}-\mathcal{T}\partial_{x}h_{T}+jh_{T}, 
\label{jl}
\end{equation}
and the charge current is equal to: 
\begin{equation}
j_{T}=D_{T}\partial_{x}h_{T}+\mathcal{T}\partial_{x}h_{L}+jh_{L}. 
\label{jt}
\end{equation}
The transport coefficients, which enter the definitions of the currents $%
j_{L,T},$ have the following physical meaning: $D_{L,T}$ are diffusion
coefficients for energy and charge excitations, $\mathcal{T}$ $\ $is
responsible for conversion of charge current to energy current and vice
versa, while $N$ plays the role of the DOS of electron excitations. Finally, 
$j$ is determined by the spectral supercurrent $J$, see Eq.(\ref{jra}).
These quantities are modified compared to their equilibrium values as a
result of the time dependence of the electric field, see Appendix \ref{AB}
for explicit expressions for them in terms of the unperturbed $\theta$ and $%
\phi$.

\subsection{Spectral corrections.}

Let us now turn to the corrections to the spectral functions, $\delta\hat {g}%
^{R\left( A\right) }$ (we will omit superscripts (R,A) below, since it can
not lead to any confusion). Naively, each of these two matrices in the
particle/hole space has four components:%
\begin{equation}
\delta\hat{g}\left( \epsilon,\epsilon^{\prime},x\right) =\left( 
\begin{array}{cc}
u_{pp}\left( \epsilon,x\right) & u_{ph}\left( \epsilon,x\right) \\ 
u_{hp}\left( \epsilon,x\right) & u_{hh}\left( \epsilon,x\right)%
\end{array}
\right) \delta\left( \epsilon-\epsilon^{\prime}-\omega\right) ,   \label{dg}
\end{equation}
but the normalization condition $\delta\left( \hat{g}\cdot\hat{g}\right) =%
\hat{g}\cdot\delta\hat{g}+\delta\hat{g}\cdot\hat{g}=0$ allows to express
diagonal components in terms of the off-diagonal ones:

\begin{equation}
\left( 
\begin{array}{c}
u_{pp}\left( \epsilon\right) \\ 
u_{hh}\left( \epsilon\right)%
\end{array}
\right) =\frac{\hat{M}}{1-\text{th}^{2}\frac{\theta}{2}\text{th}^{2}\frac{%
\theta_{-}}{2}}\left( 
\begin{array}{c}
u_{ph}\left( \epsilon\right) \\ 
u_{hp}\left( \epsilon\right)%
\end{array}
\right)   \label{normresol}
\end{equation}
with matrix $M$ given by:

\begin{equation}
\hat{M}=\left( 
\begin{array}{cc}
e^{i\phi_{-}}\text{th}\frac{\theta_{-}}{2}\left( 1-\text{th}^{2}\frac{\theta 
}{2}\right) & -e^{i\phi}\text{th}\frac{\theta}{2}\left( 1-\text{th}^{2}\frac{%
\theta_{-}}{2}\right) \\ 
-e^{-i\phi}\text{th}\frac{\theta}{2}\left( 1-\text{th}^{2}\frac{\theta_{-}}{2%
}\right) & e^{i\phi_{-}}\text{th}\frac{\theta_{-}}{2}\left( 1-\text{th}^{2}%
\frac{\theta}{2}\right)%
\end{array}
\right)   \label{M}
\end{equation}
and notation $f_{-}\left( \epsilon\right) =f\left( \epsilon-\omega\right) $
is used. Parametrization (\ref{normresol}) reduces the number of independent
components in $\delta\hat{g}$ to two: $u_{ph},~u_{hp}$. In the contacts,
these functions are driven by the time-dependent voltage:

\begin{gather}  \label{u1}
u_{ph}\left( \epsilon,x=0,1\right) =u_{ph,0}\left( \epsilon ,x=0,1\right) ,
\\
u_{ph,0}\left( \epsilon,x\right) =\frac{e^{i\phi}\sinh\theta+e^{i\phi_{-}}%
\sinh\theta_{-}}{\omega}e\varphi\left( x\right) .   \label{u2}
\end{gather}
Similar equations valid for $u_{hp}$ can be obtained from Eqs.(\ref{u1},\ref%
{u2}) by the replacement $\phi\rightarrow-\phi$. In the wire, functions $%
u_{ph}(x)$ and $~u_{hp}(x)$ are determined by the conservation laws of the
spectral currents, which take the following form:

\begin{gather}
\partial_{x}j_{ph}+\left( 2i\epsilon-i\omega-\tau_{in}^{-1}\right) u_{ph}+
\label{jss} \\
+ie\varphi\left[ e^{i\phi}\sinh\theta-e^{i\phi_{-}}\sinh\theta_{-}\right] =0.
\end{gather}
Similar equation is valid for $u_{hp}$ and $j_{hp}$ with substitution $%
\phi\rightarrow-\phi$. The spectral currents read:%
\begin{equation}
\left( 
\begin{array}{c}
j_{ph} \\ 
j_{hp}%
\end{array}
\right) =\hat{D}_{S}\left( 
\begin{array}{c}
\partial_{x}u_{ph} \\ 
\partial_{x}u_{hp}%
\end{array}
\right) +\hat{J}_{S}\left( 
\begin{array}{c}
u_{ph} \\ 
u_{hp}%
\end{array}
\right) \text{,}   \label{spec}
\end{equation}
where%
\begin{equation}
\hat{D}_{S}=\left( 
\begin{array}{cc}
D_{S} & e^{i\left( \phi+\phi_{-}\right) }\bar{D}_{S} \\ 
e^{-i\left( \phi+\phi_{-}\right) }\bar{D}_{S} & D_{S}%
\end{array}
\right)   \label{DS}
\end{equation}
and%
\begin{equation}
\hat{J}_{S}=\left( 
\begin{array}{cc}
J_{S}+\frac{J+J_{-}}{\cosh\theta+\cosh\theta_{-}} & e^{i\left( \phi+\phi
_{-}\right) }\bar{J}_{S} \\ 
e^{-i\left( \phi+\phi_{-}\right) }\bar{J}_{S} & J_{S}-\frac{J+J_{-}}{%
\cosh\theta+\cosh\theta_{-}}%
\end{array}
\right) .   \label{JS}
\end{equation}
The spectral transport coefficients $D_{S},\bar{D}_{S}$ and $J_{S},\bar{J}%
_{S}$ which enter these expressions, are provided in the Appendix \ref{AB}.

\section{Results}

\begin{figure}[ptb]
\centering
\includegraphics[width=230pt]{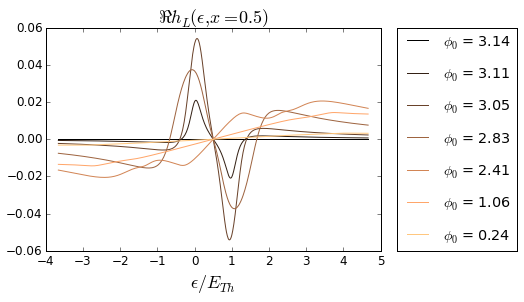}
\caption{Variation of the longitudinal distribution function at $T=15E_{Th}$%
, $\hbar\protect\omega=E_{Th}$ with varying the phase difference (color
online).}
\label{fig:hl}
\end{figure}

\begin{figure}[ptb]
\centering
\includegraphics[width=230pt]{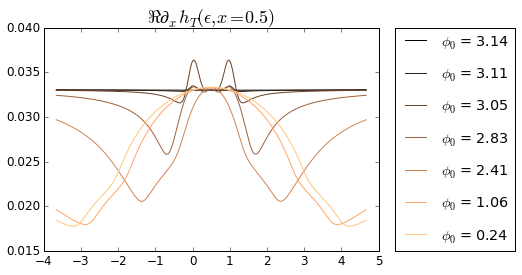}
\caption{Variation of the transversal distribution function at $T=15E_{Th}$, 
$\hbar\protect\omega=E_{Th}$ with varying the phase difference (color
online).}
\label{fig:ht}
\end{figure}

We start our presentation from the exemplary results for the distribution
functions, which are shown on the Figs \ref{fig:hl}, \ref{fig:ht}. In all
the figures, for inelastic rate we have assumed $\hbar\tau_{in}^{-1}=T/g$
for definiteness (value of $g$ is given in the Figure captions).

Below, we discuss the results for the admittance $Y/Y_{N}$. For comparison
with experiment, keep in mind, that dimensionless susceptibility to the
oscillating phase, introduced in Eq. (\ref{cdim}), reads $\chi/\chi_{0}=i%
\frac{\hbar\omega}{E_{Th}}Y/Y_{N}$. For the SNS junction of the experiment,
mentioned in the introduction, $\chi_{0}\approx35\mu A/\Phi_{0}$.

\subsection{High temperature.}

\begin{figure}[ptb]
\centering
\includegraphics[width=140pt]{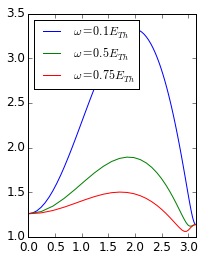}
\caption{Variation of $\mbox{Re} Y(\protect\phi_{0})/Y_{N}$ with frequency
at $T = 15E_{Th}$, $g = 40$, low frequencies (color online).}
\label{fig:loww_hight}
\end{figure}

Let us discuss the variation of the dissipative part of the admittance with
frequency at high temperature, see Figs. \ref{fig:loww_hight}, \ref%
{fig:highw_hight}. As expected, at low frequency, Lempitskii's result is
reproduced, see the curve corresponding to $\omega=0.1E_{Th}$. With growth
of the frequency, the shape of the curve drastically changes and the
dissipative part of $Y$ acquires a peak at $\phi_{0}=\pi,$ which becomes
more prominent with growth of frequency and can be clearly seen up to the
largest temperature of $T=15E_{Th}$.

\begin{figure}[ptb]
\centering
\includegraphics[width=140pt]{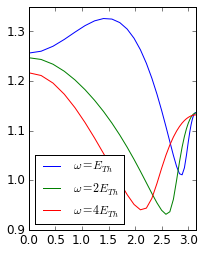}
\caption{Variation of $\mbox{Re} Y(\protect\phi_{0})/Y_{N}$ with frequency
at $T = 15E_{Th}$, $g = 40$, high frequencies (color online).}
\label{fig:highw_hight}
\end{figure}

\begin{figure}[ptb]
\centering
\includegraphics[width=140pt]{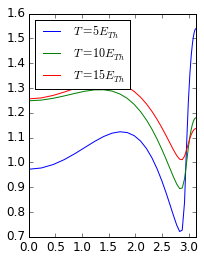}
\caption{Variation of $\mbox{Re} Y(\protect\phi_{0})/Y_{N}$ with temperature
at $\hbar\protect\omega= E_{Th}$, $g = 40$ (color online).}
\label{fig:tdep}
\end{figure}

The same kind of evolution of $\mbox{Re} Y(\phi_{0})/Y_{N}$ is shown for
different temperatures at fixed frequency $\hbar\omega=E_{Th}$ in Fig.\ref%
{fig:tdep}. Note strong peak near phase equal to $\pi$ at $T=5E_{Th}$.

In order to understand this result, recall how Eq. (\ref{lemp}) was derived.
First, we note that at $T\gg E_{Th}$ the contribution of the spectral
corrections $\delta\hat{g}$ to the electric current can be neglected and
only corrections to the distribution function ($h_{L,T}$) are important. At
finite voltage, the charge excitations, described by $h_{T},$ which enter
the wire from the superconductor and get converted into energy excitations
there, see the last term in Eq. (\ref{jl}). In the limit of $%
\hbar\omega/\Delta\ll1$ the energy excitations (described by $%
h_{L}(\epsilon,x)$) are locked in between the superconducting contacts,
since the corresponding density of states vanishes at the superconductors.
Because of that, relatively large and almost spatially independent
non-equilibrium correction to the longitudinal distribution function $%
h_{L}^{\left( \text{Lemp.}\right) }(\epsilon) = h_{L}(\epsilon)-h(\epsilon)$
in the wire is established:

\begin{equation}
h_{L}^{\left( \text{Lemp.}\right) }\approx\frac{j}{\left\langle
N\right\rangle }\frac{eV}{i\omega+\tau_{in}^{-1}}h^{\prime}\left(
\epsilon\right) \text{,}   \label{hlemp}
\end{equation}
which contributes to electric current as $I\propto\int jh_{L}^{\left( \text{%
Lemp.}\right) }d\epsilon$, leading to Eq. (\ref{lemp}). Note that this
equation seems to be inconsistent with the boundary condition, Eq. (\ref{bcl}%
). In fact, in the limit $E_{Th}\ll\Delta$ true distribution function
differs from $h_{L}^{\left( \text{Lemp.}\right) }$ in the closest vicinity
of the boundary, where it exibits large spatial gradient and sharply varies
from $h_{L}=0$ in the superconductor to $h_{L}=h_{L}^{\left( \text{Lemp.}%
\right) }$ in the wire. As a consequence, the limit of $\Delta\rightarrow%
\infty$ is singular: $h_{L}$ has a jump at $x=0$. Expanding the KE\ in the
vicinity of the contact $x=0$ we find that Eq. (\ref{bcl}) is replaced by an
effective boundary condition:%
\begin{equation}
\left. \partial_{x}h_{L}\left( \epsilon\right) \right\vert _{x=0}=\frac{%
\omega}{\zeta\left( \epsilon\right) }\left. h_{L}\left( \epsilon\right)
\right\vert _{x=0},   \label{ebc}
\end{equation}
where $\zeta\left( \epsilon\right) =$ $\left.
\partial_{x}\theta_{-}^{A}\right\vert _{x=0}-\left.
\partial_{x}\theta^{R}\right\vert _{x=0}$.

It is important that at low frequency $h_{L}$ is limited only by inelastic
processes: in the limit of $\tau_{in}\rightarrow\infty$ one has $h_{L}\propto%
\frac{V}{T}\frac{E_{Th}}{\hbar\omega}$. This is why at lowest frequencies
the correction to $h_{L}(\epsilon)$ leads to the whole effect dominated by
the Lempitskii's contribution. The properties of the transversal
distribution function $h_{T}$ are quite different. It describes charge
excitations which are free to leave the wire via Andreev reflection, so that
corrections to $h_{T}$ are relatively small at the lowest frequencies: $%
h_{T}(\epsilon)\propto\frac{V}{T}$. However, it is clear that at $\hbar
\omega\sim E_{Th}$ charged excitations described by $h_{T}(\epsilon)$ can
provide an important contribution to electric current, comparable to that
due to excitations of Andreev pairs (described by $h_{L}(\epsilon)$). If one
is interested in dissipative part of $Y,$ the corresponding condition is
even more stringent, since the real part of $h_{L}$ starts to decay already
at $\hbar\omega\sim\hbar\tau_{in}^{-1}\ll E_{Th}$.

\subsection{Low temperature.}

\begin{figure}[ptb]
\centering
\includegraphics[width=140pt]{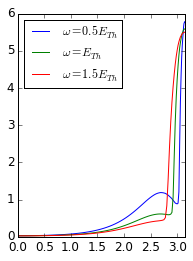}
\caption{Variation of $\mbox{Re} Y(\protect\phi_{0})/Y_{N}$ with frequency
at $T = E_{Th}$, $g = 40$ (color online).}
\label{fig:wdep_lowt}
\end{figure}

At low temperature, the dissipation is noticeable only in the vicinity of
the minigap closing, see Fig. \ref{fig:wdep_lowt}. These results are very
natural. Indeed, at $T=0$ dissipation is non-vanishing only as long as
frequency is large enough compared to the minigap $E_{g}$, in particular, at 
$\omega=0$ one has $\mbox{Re} Y\propto\delta\left( \phi_{0}-\pi\right) $.
This peak becomes broadens at finite temperature: $\delta\phi\propto\hbar%
\omega,T$. In addition, at finite $\omega$ it acquires additional structure:
observe a kink of the dissipation as $\phi_{0}$ departures from $\pi$. The
position of this kink is determined by the condition $2E_{g}\left(
\phi_{k}\right) =\omega$. Indeed, for $\phi_{k}\sim\pi,$ one has\cite%
{ivanov02}: $E_{g}\left( \phi\right) \approx\frac{\pi^{3}}{4}%
E_{Th}\left\vert 1-\phi_{\phi}/\pi\right\vert $, which gives for $%
\hbar\omega=0.5E_{Th}$: $\phi_{k}\approx3.04$. It can also be followed how
this kink shifts with growth of the frequency. At larger temperature it
becomes smoothened away, see for example the evolution of the curve on the
Fig. (\ref{fig:lowfreq}) from $T=2E_{Th}$ to $T=5E_{Th}$.

\subsection{Low frequency.}

\begin{figure}[ptb]
\centering
\includegraphics[width=140pt]{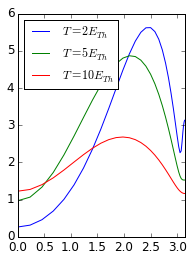}
\caption{Variation of $\mbox{Re} Y(\protect\phi_{0})/Y_{N}$ with temperature
at $\protect\omega=0.3 E_{Th}$, $g = 20$ (color online).}
\label{fig:lowfreq}
\end{figure}

Another interesting crossover in the shape of $\mbox{Re} Y(\phi_{0})/Y_{N}$
is seen at low frequencies upon variation of the temperature. It is
illustrated on the Fig. \ref{fig:lowfreq}. At moderately low temperature $%
T=2E_{Th}$ strong peak of dissipation is found at the phase difference $%
\phi_{0}\approx0.75\pi$; with temperature increase, this peak becomes more
rounded and shifts further away from $\pi$, so that curve becomes more and
more similar to Lempitskii's function $Q_{0}\left( \phi\right) $.

\section{Conclusions}

We have developed a fully microscopic approach to the calculation of a
non-stationary \textit{ac} linear response function of a S-N-S junction
under the \textit{dc} phase bias, that is valid at arbitrary relations
between temperature $T$, Thouless energy $E_{Th}$ and frequency $\omega$. We
assumed energy gap in the S terminals $\Delta$ to be much larger than all
these energy scales and took into account inelastic relaxation rate $%
\hbar/\tau_{in}\leq E_{Th}$. The shape of the dissipative response $\mbox{Re}
Y(\phi_{0})/Y_{N}$ is shown to be very sensitive to the relations between $T$%
, $E_{Th}$, $\hbar\omega$ and $\hbar/\tau_{in}$. Explicit results for the
function $\mbox{Re} Y(\phi_{0})/Y_{N}$ can be found for any choice of the
above parameters using the published codes. In particular, we have shown
that accurate solution reproduces many of the qualitative features of the
experimental results\cite{dassonneville14}, including peak at the phase
difference equal to $\pi$ at high frequencies and high temperature; we
interpret this peak as the result of charge imbalance induced by
high-frequency electric field. Still some quantitative disagreement exists:
the experimental value of dissipation at $\phi_{0}=\pi$ is higher (at the
same values of $T$ and $\omega$) than our computations provide. Possible
source of this disagreement may be related with non-zero resistance of S-N
interfaces which we did not took into account in the present calculations,
since we assume interfaces to be perfectly transmitting. It is a
straightforward task to include non-zero interface resistance into the
calculational scheme developed.

\begin{figure}[ptb]
\centering
\includegraphics[width=140pt]{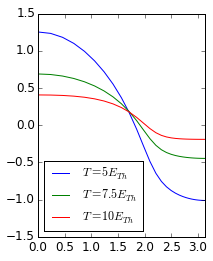}
\caption{Variation of $\mbox{Im} Y(\protect\phi_{0})/Y_{N}$ with temperature
at $\protect\omega=5 E_{Th}$, $g = 20$ (color online).}
\label{fig:imy}
\end{figure}

In our discussion, we did not touch the issue of the non-dissipative part of 
$Y\left( \phi_{0}\right) ,$ which, at moderate frequency, seems to be
reasonably well described in the Lempitskii's approximation. What lies
outside this approximation is an interesting feature at high frequency at $%
\phi _{0}\sim\pi,$ which is observed in the experiment\cite{dassonneville14}%
, as Fig. 6.21 of this reference shows. In our model, we obtain the
flattening \ of $\mbox{Im}Y$ at $\phi_{0}\sim\pi$ at $\omega\gtrsim E_{Th}$.
For example, see Fig. \ref{fig:imy} for the results at $\omega=5E_{Th}$,
which at high temperature are rather close to the experiment. However, we do
not see qualitative change of behaviour with lowering the temperature and do
not get the large drop at $\phi_{0}=\pi$ which is observed in experiment.
The nature of this drop is a very interesting problem for the future study.
Another interesting problem is to include more realistic description of
electron-electron interaction into the linear response calculation. It can
be as interesting as important due to the specific spectral properties of
the electrons, confined between superconducting reservoirs and great
sensitivity of the admittance to inelastic processes in the experimentally
relevant regime of frequency and temperature.

\emph{Acknowledgements.} The authors gratefully acknowledge H. Bouchiat, B.
Dassonneville, S. Gueron for careful reading of the manuscript and useful
discussions and P. Virtanen for comments. KT thanks the members of the
Institut f{\"{u}}r Theorie der Kondensierten Materie at KIT, where this work
was started, for their kind hospitality. KT was supported by the Paul and
Tina Gardner fund for Weizmann-TAMU collaboration and RFBR grant 13-02-00963.

\begin{appendix}
\section{Ricatti parametrization of the GF}
\label{AA}
For numerical solution of unperturbed Usadel equation, it is more
convenient to use Ricatti parametrization:
\begin{equation}
\hat{G}_{0}^{R}=\frac{1}{1-ab}\left(
\begin{array}
[c]{cc}%
1+ab & 2a\\
-2b & -1-ab
\end{array}
\right).
\end{equation}
In this parametrization, spectral Usadel equation reads:
\begin{equation}
Da^{\prime\prime}+2i\epsilon a=-\frac{2Dba^{\prime2}}{1-ab},~Db^{\prime\prime
}+2i\epsilon b=-\frac{2Dab^{\prime2}}{1-ab}.
\end{equation}
In the main part, we hold to trigonometric parametrization, see Eq
(\ref{g0}), which makes formulae more compact. The relationship between the two parametrizations
is as follows:
\begin{equation}
a=e^{i\phi }\text{th}\frac{\theta }{2},~b=e^{-i\phi }\text{th}\frac{\theta }{2}.
\end{equation}
\section{Transport coefficients at finite frequency}
\label{AB} Here we present expressions for transport coefficients at non-zero
frequency, which enter Eqs (\ref{kl}), (\ref{kt}). Energy/charge diffusion coefficients read:
\begin{equation}
D_{L,T}=1-\cosh\theta^{R}\cosh\theta_{-}^{A}\pm\cos\left(  \phi_{-}^{A}%
-\phi^{R}\right)  \sinh\theta^{R}\sinh\theta_{-}^{A},
\end{equation}
anomalous transport coefficient:
\begin{equation}
\mathcal{T}=-i\sin\left(  \phi_{-}^{A}-\phi^{R}\right)  \sinh\theta^{R}%
\sinh\theta_{-}^{A},
\end{equation}
and density of states:
\begin{equation}
N=\cosh\theta^{R}-\cosh\theta_{-}^{A}.
\end{equation}
Finally, the spectral supercurrent reads:
\begin{equation}
\label{jra}
j=J^{R}-J_{-}^{A}%
\end{equation}
The frequency of oscillations enters these expression by the energy shifts, which
are shown by\ the following notation: $f_{\pm}\left(  \epsilon\right)
=f\left(  \epsilon\pm\omega\right)  $.
The spectral transport coefficients, which enter Eq (\ref{jss}) read:
\begin{gather}
D_{S}=\frac{1+\cosh \theta \cosh \theta _{-}}{\cosh \theta +\cosh \theta _{-}%
},\\
~\bar{D}_{S}=\frac{\sinh \theta \sinh \theta _{-}}{\cosh \theta +\cosh
\theta _{-}}
\end{gather}
and
\begin{gather}
J_{S}=\left( 1+\cosh \theta \cosh \theta _{-}\right) \partial _{x}\frac{1}{%
\cosh \theta +\cosh \theta _{-}}, \\
~\bar{J}_{S}=\sinh \theta \sinh \theta _{-}\partial _{x}\frac{1}{\cosh
\theta +\cosh \theta _{-}}\text{.}
\end{gather}
\end{appendix}

\bibliography{sns}

\end{document}